\documentclass[aps, prb, amsmath, amssymb, 
superscriptaddress, 
twocolumn,
citeautoscript]{revtex4-1}
\usepackage{graphicx}
\usepackage{hyperref}
\usepackage{textcomp}
\usepackage{color, soul, textcomp}

\hypersetup{colorlinks=true, urlcolor=blue, citecolor=blue, linkcolor=blue, anchorcolor=blue}  

\newcommand{\dfr}{\ensuremath{\Delta f_{\rm rep}\ }}
\newcommand{\Sin}{Si$_3$N$_4\ $}

\begin{document}
\title{On-chip dual comb source for spectroscopy}

\author{Avik Dutt}
\affiliation{School of Electrical and Computer Engineering, Cornell University, Ithaca, New York 14853, USA}
\affiliation{Department of Electrical Engineering, Columbia University, New York, New York 10027, USA}

\author{Chaitanya Joshi}
\affiliation{School of Applied and Engineering Physics, Cornell University, Ithaca, New York 14853, USA}
\affiliation{Department of Applied Physics and Applied Mathematics, Columbia University, New York, New York 10027, USA}

\author{Xingchen Ji}
\affiliation{School of Electrical and Computer Engineering, Cornell University, Ithaca, New York 14853, USA}
\affiliation{Department of Electrical Engineering, Columbia University, New York, New York 10027, USA}

\author{Jaime Cardenas}
\email[Currently at The Institute of Optics, University of Rochester, Rochester, New York 14627, USA]{}
\affiliation{Department of Electrical Engineering, Columbia University, New York, New York 10027, USA}

\author{Yoshitomo Okawachi}
\affiliation{Department of Applied Physics and Applied Mathematics, Columbia University, New York, New York 10027, USA}

\author{Kevin Luke}
\affiliation{School of Electrical and Computer Engineering, Cornell University, Ithaca, New York 14853, USA}

\author{Alexander L. Gaeta}
\affiliation{Department of Applied Physics and Applied Mathematics, Columbia University, New York, New York 10027, USA}

\author{Michal Lipson}
\email[email: ]{ad654@cornell.edu, ml3745@columbia.edu}
\affiliation{Department of Electrical Engineering, Columbia University, New York, New York 10027, USA}

\begin{abstract}
Dual-comb spectroscopy is a powerful technique for real-time, broadband optical sampling of molecular spectra which requires no moving components. Recent developments with microresonator-based platforms have enabled frequency combs at the chip scale. However, the need to precisely match the resonance wavelengths of distinct high-quality-factor microcavities has hindered the development of an on-chip dual comb source. Here, we report the first simultaneous generation of two microresonator combs on the same chip from a single laser. The combs span a broad bandwidth of 51 THz around a wavelength of 1.56 \textmu m. We demonstrate low-noise operation of both frequency combs by deterministically tuning into soliton mode-locked states using integrated microheaters, resulting in narrow ($<$ 10 kHz) microwave beatnotes. We further use one mode-locked comb as a reference to probe the formation dynamics of the other comb, thus introducing a technique to investigate comb evolution without auxiliary lasers or microwave oscillators. We demonstrate broadband high-SNR absorption spectroscopy of dichloromethane spanning 170 nm using the dual comb source over a 20 \textmu s acquisition time. Our device paves the way for compact and robust dual-comb spectrometers at nanosecond timescales.
\end{abstract}

\maketitle

\section{Introduction}

Optical frequency combs \cite{cundiff_colloquium_2003} are a revolutionary technology with widespread applications in precision spectroscopy \cite{hansch_nobel_2006}, microwave signal synthesis \cite{huang_spectral_2008, fortier_generation_2011}, frequency metrology \cite{hall_nobel_2006}, optical clocks, communications \cite{wang_observation_2012, pfeifle_coherent_2014, pfeifle_optimally_2015} and astronomical spectrograph calibration \cite{li_laser_2008, steinmetz_laser_2008, phillips_calibration_2012, ycas_demonstration_2012, zajnulina_generation_2015}. A frequency comb consists of a sequence of hundreds or thousands of narrow, discrete spectral lines equidistant in frequency, separated by the repetition rate $f_{\rm rep}$. Many of the previous demonstrations of frequency comb sources have utilized femtosecond mode-locked lasers \cite{diddams_direct_2000, jones_carrier-envelope_2000, diddams_evolving_2010}. In recent years, there has been significant development of chip-scale comb sources based on the Kerr nonlinearity using high quality factor (Q) microresonators \cite{kippenberg_microresonator-based_2011, matsko_optical_2005}. The past decade has seen the development of numerous platforms for generating such microresonator-based frequency combs, including silica\cite{delhaye_optical_2007, agha_four-wave-mixing_2007}, silicon nitride (Si$_3$N$_4$)  \cite{levy_cmos-compatible_2010, foster_silicon-based_2011}, crystalline fluorides \cite{savchenkov_tunable_2008, henriet_kerr_2015, wang_mid-infrared_2013}, Hydex glass \cite{razzari_cmos-compatible_2010}, aluminum nitride \cite{jung_optical_2013}, diamond \cite{hausmann_diamond_2014}, silicon \cite{griffith_silicon-chip_2015} and AlGaAs \cite{pu_efficient_2016}. These microresonators, when pumped by a monochromatic continuous wave (cw) laser, generate frequency sidebands through four-wave mixing (FWM) parametric oscillation and undergo nontrivial nonlinear dynamics to produce a comb of frequencies that can span an octave of bandwidth \cite{delhaye_octave_2011, okawachi_octave-spanning_2011}.

Dual comb spectroscopy (DCS) enables broadband optical sampling of molecular absorption spectra with high signal-to-noise ratio (SNR) and tens-of-\textmu s acquisition times, without any moving parts, through use of the high coherence of the comb lines \cite{schiller_spectrometry_2002, keilmann_time-domain_2004, bernhardt_cavity-enhanced_2010, coddington_coherent_2008, coddington_coherent_2010, coddington_dual-comb_2016, ideguchi_adaptive_2014}. This technique utilizes two frequency combs with slightly different repetition rates, $f_{\rm rep1}$ and $f_{\rm rep2} = f_{\rm rep1} + \dfr$. The heterodyne beating between these two combs generates a sequence of beat notes in the radio-frequency (RF) domain, spaced by the difference in repetition rates \dfr, ensuring a one-to-one mapping of the optical comb lines to the RF beat notes. This beating down-converts the spectral information spanning tens to hundreds of THz in the optical domain to a few GHz in the RF domain [Fig. 1]. DCS has been traditionally implemented using stabilized \cite{coddington_coherent_2008, coddington_coherent_2010} as well as free running \cite{ideguchi_adaptive_2014, mehravar_real-time_2016} mode-locked Ti:sapphire or fibre lasers and therefore often requires bulky tabletop setups. Chip-based frequency combs have the potential to implement DCS in a compact and robust integrated platform.
	
The challenge in performing DCS using an integrated platform lies in the requirement for mode-locking each of the combs generated in both microcavities and aligning their resonance wavelengths to a single external pump laser wavelength (in order to ensure high relative coherence between the two spectral combs). Two experiments have demonstrated on-chip integration of dual combs using semiconductor disk lasers \cite{link_dual-comb_2015} and quantum cascade lasers \cite{villares_-chip_2015}, with narrow optical bandwidths of  0.08 THz and 0.96 THz, respectively. The development of a dual-comb source based on Kerr nonlinear microresonators, used in combination with dispersion engineered waveguides, has the potential to increase the bandwidth by orders of magnitude without compromising the device footprint or the on-chip integration capability. Recently, Suh \emph{et al}.\cite{suh_microresonator_2016} demonstrated dual combs based on wedge resonators using two separate chips pumped by two independent lasers, which are limited in their optical bandwidth due to dispersion, and in their achievable relative coherence time without using active locking of the two lasers. Similarly, Yu \emph{et al}. \cite{yu_silicon-chip-based_2016} reported mid-infrared DCS using two separate chips pumped by an optical parametric oscillator.
	
In this work, we present for the first time the generation of dual Kerr frequency combs on a single chip, using a single pump laser and dispersion-engineered Si$_3$N$_4$ microrings to produce wide optical bandwidths and long coherence times. We demonstrate soliton modelocking of both combs using integrated thermal tuning, leading to narrow microwave beat notes with linewidths $<$ 10 kHz and sech$^2$ –shaped optical spectra spanning 51 THz with minimal mode crossings. Integrated thermal tuning based on platinum microheaters also allows us to use a low-noise non-tunable laser as the pump, resulting in a long mutual coherence time of 100 \textmu s even without active feedback and stabilization. Furthermore, we use one of the soliton mode-locked combs as a reference and investigate the generation dynamics of the second comb as the latter is tuned into the mode-locked state by measuring the beat note between the combs, thus illustrating the power of the dual-comb heterodyne technique to study the formation dynamics of frequency combs. Finally, we perform absorption spectroscopy of dichloromethane in the near-infrared spanning a 170 nm wavelength range, over a fast acquisition time of 20 \textmu s without any averaging, using the demonstrated dual-comb source.

\section{Results}
\subsection{Device design}

The dual comb is generated using cascaded \Sin microring resonators pumped with a single cw laser. The cascaded rings (R1 and R2) have slightly different nominal radii of 50.04 \textmu m and 49.98 \textmu m respectively [Fig.~\ref{fig:1_schematic}(a)], and are coupled to the same bus waveguide. \Sin has proven to be a versatile platform for on-chip photonics due to its low linear loss ($\sim$ 0.8 dB m$^{-1}$) \cite{shah_hosseini_high_2009, tien_ultra-high_2011, luke_overcoming_2013, li_vertical_2013, spencer_integrated_2014, ji_breaking_2016}, high refractive index ($\sim 2$), negligible nonlinear absorption in the near-infrared, and compatibility with complementary metal-oxide-semiconductor (CMOS) fabrication processes\cite{moss_new_2013}. \Sin waveguides are particularly attractive for integrated nonlinear photonics because of the high nonlinear index ($n_2 \sim 2\times 10^{-19}$ m$^2$ W$^{-1}$) \cite{levy_cmos-compatible_2010, kruckel_linear_2015}, high confinement \cite{kruckel_linear_2015, gondarenko_high_2009, epping_high_2015}, and the ability to tailor the group velocity dispersion (GVD) by engineering the waveguide geometry \cite{turner_tailored_2006, chavez_boggio_dispersion_2014}. In our device, the waveguides have a height of 730 nm and a width of 1500 nm to ensure low anomalous GVD around the pump wavelength of 1561 nm, which is necessary for broadband comb generation \cite{kippenberg_microresonator-based_2011, delhaye_octave_2011, okawachi_octave-spanning_2011}. The coupling gap between the waveguide and the resonator is 350 nm.

We overcome the challenge of aligning the resonances of both high Q rings (Q $>$ 600,000) to the pump wavelength by fabricating integrated platinum microheaters on top of the \Sin rings to tune the resonance wavelength of the rings using the thermo-optic effect. This is in contrast to what has been done in most previous experiments with Kerr combs, where the laser wavelength is tuned to a cavity mode of the microresonator \cite{herr_temporal_2013}. The microheaters have highly localized heat flow, reduced thermal crosstalk \cite{sherwood-droz_optical_2008} and fast response times, and enable us to stably and repeatedly tune the rings into resonance with the pump \cite{zeng_design_2014, miller_tunable_2015, xue_normal-dispersion_2015, xue_thermal_2016, joshi_thermally_2016}. The rise and fall times of the microheaters are 24 \textmu s and 22 \textmu s respectively (See Supplementary information fig. S1), which allow us to tune into resonance at speeds of several tens of kHz \cite{joshi_thermally_2016}; this is significantly faster than using Peltier elements to tune the temperature of the entire chip. Details of the fabrication process can be found in the Methods section.

\subsection{Dual comb optical spectra}

The heaters enable tuning of either one or both combs into resonance to generate frequency comb spectra spanning 51 THz (400 nm) near the pump wavelength of 1561.4 nm, as shown in Fig. \ref{fig:2_optical}(a)-(c). We excite ring R1 with $\sim$ 105 mW of power in the bus waveguide and tune its heater without tuning the heater on R2 to generate the spectrum shown in Fig. \ref{fig:2_optical}(a). Similarly, Fig. \ref{fig:2_optical}(b) shows the spectrum obtained by tuning the heater on R2 without tuning the heater on R1. These individual comb spectra allow us to characterize the mode spacings of the combs to within the 0.01 nm resolution of the optical spectrum analyzer (OSA). Both rings generate a spectrum spanning 400 nm (51 THz). We infer a mode spacing $f_{\rm rep1} = 451.4\pm 0.4$ GHz for the R1 comb and $f_{\rm rep2} = 452.8\pm 0.4$ GHz for the R2 comb. The difference between the mode spacings of the combs is $\dfr = 1.4 \pm 0.8$ GHz, which is well within the bandwidth of commercially available fast photodiodes, whereas the repetition rate frequency of each individual comb is much larger than what can be directly detected with fast photodiodes. This is true in general for microresonator combs, which have repetition rates from tens to hundreds of GHz. Thus, the dual comb technique frequency downconverts the repetition rate of microresonator combs making it amenable to study the repetition rate beat note using fast photodiodes. Note that the large error bar in \dfr is limited by the resolution of the OSA, and RF heterodyne beating measurements shown later determine \dfr to a much higher accuracy. Fig. \ref{fig:2_optical}(c) shows the spectrum of the dual comb generated when both R1 and R2 are on resonance with the laser. As can be seen from the inset of Fig. \ref{fig:2_optical}(c), pairs of closely spaced modes are clearly distinguishable, indicating a dual comb spectrum. It is worth noting that \dfr is much larger than the linewidth of each cavity mode (230 MHz for R1, 310 MHz for R2), so that the comb lines generated in R1 are not resonant with the cavity modes of R2, except for the pump. This ensures that the comb lines from R1 do not couple into the R2 resonances and interact with the comb lines in R2, despite the cascaded geometry (see Supplementary information section II).

By engineering the thermal tuning of the heaters with respect to the comb, we independently tune the two rings at speeds of tens of kHz and achieve modelocking of the two spectral combs using a fixed-frequency pump laser. The combs generated in R1 and R2 exhibit single-soliton modelocking with sech$^2$-spectral envelopes without significant mode crossings. We tune into the low-noise mode-locked states of both the combs by applying a voltage ramp to the microheaters of rings R1 and R2 over a period of 50 \textmu s using an arbitrary waveform generator, corresponding to a scan speed of 20 kHz as shown in Fig. \ref{fig:2_optical}(e), and as detailed in Ref. \onlinecite{joshi_thermally_2016}. The downward voltage ramp is applied on a ring when the pump laser is initially on the blue-detuned side of the ring resonance which blueshifts the ring resonance towards the pump laser. A final increase in voltage redshifts the resonance, providing us deterministic access to the soliton states. When the pump wavelength is on the red-detuned side of the resonance, we observe a series of abrupt steps in the pump transmission concomitant to the formation of multi-soliton states with different numbers of solitons propagating per roundtrip around the microring, similar to what has been routinely observed in recent work on soliton modelocking\cite{herr_temporal_2013, joshi_thermally_2016, yi_soliton_2015}. By controlling the magnitude of the final upward voltage ramp, one can control the number of solitons present in the comb. For example, Fig. \ref{fig:2_optical}(d) depicts a spectrum with a single soliton state in R1 and a two-soliton harmonic mode-locked state in R2. Note that recent experiments have demonstrated soliton modelocking in single microresonators using pump wavelength tuning\cite{herr_temporal_2013}, two-step pump power control\cite{yi_soliton_2015, brasch_bringing_2016} and integrated thermal tuning\cite{joshi_thermally_2016}. Here we choose thermal tuning since it allows us to use a fixed-frequency non-tunable low-noise laser\cite{numata_performance_2010} (RIO Orion, linewidth $<$ 3kHz) as the pump and enables independent tuning of the two rings at sufficiently high speeds of tens of kHz. Tunable external cavity diode lasers, conventionally used for Kerr frequency comb generation, have a broader linewidth ($>$ 100 kHz), have higher phase noise and frequency jitter, and are typically not amenable to tuning at the required speeds. It is important to note that the soliton states, once generated, stably exist for hours without any active feedback, in spite of the drifts in fiber-to-chip coupling and pump power.

\subsection{Multi-heterodyne beat notes}
We observe multi-heterodyne beat notes between the combs when both rings are tuned into resonance with the pump laser to generate dual combs. For this measurement, the output of the chip is sent to a wavelength-division multiplexing (WDM) filter to remove the pump. Since the comb lines are symmetric in frequency relative to the pump, the beat notes of the comb lines on the red side of the pump interfere with the beat notes on the blue detuned side. To avoid this and hence enable one-to-one mapping of the RF beat notes to the optical wavelength, we use short and long pass filters to characterize the beat notes on the shorter and longer wavelength sides of the dual comb (see Supplementary Information section II). The filtered comb is sent to a 45 GHz photodiode (New Focus 1014) followed by a three-stage 40 GHz RF amplifier (Centellax OA4MVM3), and the resulting RF signal is monitored on a real time oscilloscope (LeCroy LabMaster 10 Zi-A, 8-bit resolution, 80 GSa/s). In Fig. \ref{fig:3_RF}(a) we plot the time-domain interferogram over 20 \textmu s. Fig. \ref{fig:3_RF}(b) shows the same interferogram expanded 2000 times, and a periodic signal is clearly visible repeating every 900 ps (=1/\dfr). To further elucidate the details of the time domain interferogram, we amplify the filtered dual comb using an L-band EDFA and plot the amplified and expanded version of the interferogram in Fig. \ref{fig:3_RF}(c). A fast Fourier transform (FFT) of the time domain signal yields 27 RF beat notes spaced by 1.12 GHz, which corresponds to the difference in mode spacings of the two combs [Fig. \ref{fig:3_RF}(d)]. The number of beat notes detected is limited by the noise floor of the FFT and the decreasing power in the comb lines away from the pump. The beat notes below 10 GHz have very high SNR in excess of 40 dB. The center frequency of the first beat note is 1.12 GHz, which is commensurate with the difference in mode spacings estimated from the OSA scans within the uncertainty of the OSA measurement. It also matches reasonably well with the difference in mode spacings calculated based on the radii of R1 and R2 (see Supplementary Information section II).

The soliton mode-locked dual comb exhibits low-noise RF beat notes with a narrow 3-dB linewidth of 10 kHz [Fig. \ref{fig:3_RF}(c)], which ensures a long relative coherence time [$\tau_{\rm rel} = (10  {\rm kHz})^{-1} = 100$ \textmu s]. The first beat note has an SNR exceeding 60 dB over a span of 2 MHz with a sweep time of 500 \textmu s. Our high SNR over fast acquisition times is enabled by (i) a short single-shot acquisition time (limited by $\tau_{\rm acq, min} = 2/\dfr = 1.79$ ns to resolve individual beat notes), and (ii) our ability to perform complete spectral acquisitions well within the relative coherence time $\tau_{\rm rel} = 100$ \textmu s. We choose an acquisition time $\tau_{\rm rel} = 20$ \textmu s to satisfy the condition $\tau_{\rm acq, min} < \tau_{\rm acq} < \tau_{\rm rel}$. Note that the relative coherence between the two mode-locked combs is high even though the combs are in the free-running state (i.e. there is no active feedback and stabilization) since both the combs are generated from the same low-noise pump laser. The observed beat note linewidth of 10 kHz is not fundamental but is limited by the noise in the arbitrary waveform generator used to control the heaters on the rings. Shorter linewidths ($<$ 1 kHz) at the repetition rate beat note have been reported for single combs without active feedback\cite{herr_temporal_2013, brasch_photonic_2016, li_low-pump-power_2012}, and hence it should be possible to reach even longer coherence times using a lower noise voltage source to control the heaters.

To underscore the importance of modelocking for the generation of narrow beat notes, we also study the dual comb in the high-noise state which exhibits a much broader beat note 3-dB linewidth. The linewidth of the dual comb beat note in this state is $\sim$ 300 MHz, which is rather broad for precision spectroscopy or microwave signal generation (see Supplementary Information section III). This is because in the high RF-amplitude noise state, the frequencies and the relative phases of the comb lines randomly fluctuate in time.

\subsection{Evolution of beat notes}
We show that the integrated dual comb source enables one to directly observe the comb formation dynamics without the need for a secondary laser or microwave oscillator. A microresonator frequency comb starts as primary sidebands which develop into minicombs, which further evolve into a high-noise state and finally undergoes a transition to a mode-locked state through higher-order FWM interactions\cite{herr_temporal_2013, brasch_photonic_2016, li_low-pump-power_2012}. Such dynamics have been studied experimentally by characterizing the repetition rate beat note, but that often requires auxiliary lasers or RF local oscillators for heterodyne downmixing \cite{saha_modelocking_2013, herr_universal_2012, li_low-pump-power_2012}. Here we use the dual comb heterodyne technique for the same purpose, obviating the need for an auxiliary laser. We monitor the evolution of the RF noise of the comb from R2 as it is tuned into the soliton mode-locked state, by beating it with a mode-locked comb from R1. The rings used in this measurement have a difference in radius of 1 \textmu m, corresponding to an 8.6 GHz beat note spacing. The larger beat note spacing allows us to study the noise characteristics of combs, which typically span up to several GHz in the high noise state. A voltage ramp is applied initially to the heater on R1, as described in the previous section, to generate a soliton mode-locked comb, which is then used as a reference to probe the different stages of comb formation in R2. The voltage applied to the heater on R2 is changed and the RF spectrum around the first heterodyne beat note at 8.6 GHz is simultaneously recorded. Fig. \ref{fig:4_evolution} shows the evolution of the beat note as R2 goes through several stages of comb formation, finally culminating in the narrow, high-SNR beat note characteristic of dual mode-locked combs described in the previous section. On the blue-detuned side, a single (state I) or multiple (state II) low-SNR beat notes are observed. With further tuning, the number of beat notes as well as the width of each beat note increases, eventually leading to the broad RF beat note characteristic of the high-noise state (stage III and IV). By continuing the scan, we access the red-detuned side of the resonance, resulting in the formation of a high-SNR narrow beat note (Stage V, see also Supplementary movie). Further continuous scanning in the same direction leads to the ring falling out of resonance with the laser. Note that since there is a one-to-one mapping between each pair of closely spaced lines of the dual comb and the corresponding beat note in the downmixed RF spectrum, one can in principle study the dynamics of the comb line-by-line by experimentally analyzing the desired RF heterodyne beat note.

\subsection{Dual-comb spectroscopy of dichloromethane}
We perform liquid phase absorption spectroscopy of dichloromethane near the pump wavelength of 1561 nm using the dual comb source and show that the recorded spectrum using DCS agrees very well with the corresponding spectrum obtained using an OSA and a broadband supercontinuum (Fianium SC-450-4) as a light source. Dichloromethane is a widely used organic solvent with broad absorption features in the near-IR. For spectroscopy, the output of the chip is collected with an aspheric lens and sent directly through a cuvette containing dichloromethane [Fig. \ref{fig:5_DCS}(a)]. The absorption spectrum is obtained by recording the RF heterodyne beating of the dual combs with and without the sample of dichloromethane, and it is then corroborated with the corresponding optical spectrum (see fig. \ref{fig:5_DCS}(b)). As stated earlier, we acquire a time domain interferogram over 20 \textmu s at 80 GSa/s, and perform an FFT to extract the beat note information in the RF domain. We achieved a high SNR without averaging multiple spectra, thus indicating the real-time nature of the acquisition system. The quartz cuvette showed no significant absorption in this wavelength range. Note that the optical bandwidth of 170 nm accessible by the dual comb is narrower in our experiment than the full extent of the dual comb spectrum (400 nm) due to the photodiode's sharp roll-off in responsivity beyond 1640 nm and the decreasing power of the dual comb lines in accordance with the sech$^2$-shape which restricts the SNR of wavelengths shorter than 1460 nm. 

\section{Discussion}
We have demonstrated the generation of two independent frequency combs based on \Sin microresonators with slightly different repetition rates on the same chip. The generated combs have a broadband spectrum spanning 400 nm (51 THz), high repetition rates of 450 GHz with a repetition rate difference of 1.12 GHz. Dispersion engineering by tailoring the waveguide dimensions should allow us to further broaden the bandwidth of the combs up to an octave \cite{okawachi_octave-spanning_2011, delhaye_octave_2011, brasch_photonic_2016}. The combs possess a high degree of relative coherence since they are generated from the same fixed-frequency low-noise cw pump laser, as evidenced by narrow RF heterodyne beat notes with a linewidth of 10 kHz. Both the combs are operated in the single-soliton mode-locked state with the help of integrated platinum heaters to tune the resonance wavelengths of the microresonators.

The approach we have developed can be extended to other wavelength ranges, such as the mid-IR ``molecular fingerprint" region, where several molecules have their strongest fundamental rotovibrational resonances, as shown by the recent experiments by Yu \emph{et al.} \cite{yu_silicon-chip-based_2016}. The repetition rate of our dual comb (450 GHz) provides sufficient resolution for spectroscopy of liquids and solids, which typically have broad absorption features of a few THz or more, as demonstrated in the real-time DCS measurement of dichloromethane reported here. Higher resolution spectroscopy, as required typically for gases, is possible by tuning the laser and the resonator synchronously once the comb is generated. High-repetition-rate combs such as the ones shown here are desirable for fast acquisition timescales, as the speed of acquisition scales with an increase of the comb tooth spacing $f_{\rm rep}$ \cite{coddington_dual-comb_2016}. Such combs are promising for investigating rapid dynamical processes such as chemical reactions \cite{fleisher_mid-infrared_2014} or single-shot measurements in turbulent environments \cite{coddington_dual-comb_2016}.

In the present work, the dual-comb source is coupled out of the chip using a single waveguide, which is adequate for absorption spectroscopy as it requires detecting only the amplitude information. Dispersive measurements which require phase sensitive detection \cite{schiller_spectrometry_2002} could be enabled by designing two separate outputs for the two combs, one used as the reference arm and the other sent to the sample to be measured. This would require incorporating either drop ports on both the rings or splitting the pump laser before sending them to two waveguide-coupled rings in a non-cascaded geometry. The latter configuration also permits combs with smaller difference in repetition rates ($f_{\rm rep} \sim < 10$ MHz, see Supplementary Information section II), making time-domain measurements more feasible, at the cost of increasing the minimum time needed for acquiring a single-shot spectrum. Combined with such improvements, a chip-scale microresonator based dual-comb source opens the door to realizing a compact portable spectrometer for stand-off molecular sensing in the field.

\section*{Methods}
\subsection{Device fabrication}
The devices were fabricated using a process similar to that described in Ref. \onlinecite{luke_overcoming_2013} and Ref. \onlinecite{ji_breaking_2016}. We start with a 4-inch diameter virgin Si wafer and thermally oxidize it to obtain a 4 \textmu m buried oxide layer to be used as the undercladding. Trenches are defined before nitride deposition on the oxide to mitigate stress-induced crack propagation in the film. A 730 nm thick layer of stoichiometric silicon nitride is grown using low-pressure chemical vapour deposition (LPCVD) in two steps. The waveguides and resonators are patterned with electron beam lithography on a JEOL 9500FS system using ma-N 2403 resist. An alternative procedure uses an oxide hard mask to better transfer patterns to the nitride layer from the resist. This oxide is deposited using plasma-enhanced chemical vapour deposition (PECVD). The pattern is etched using inductively coupled plasma reactive ion etching (ICP RIE) with CHF$_3$, N$_2$ and O$_2$ gases. The devices are annealed in an argon atmosphere at 1200 \textdegree C after stripping the resist to remove residual N-H bonds which introduce optical loss in the nitride film. The devices are clad with 500 nm of high temperature oxide (HTO) at 800 \textdegree C followed by a 2.5 \textmu m thick overcladding of PECVD oxide. We sputter 100 nm of platinum on top of the oxide cladding and define the heaters using photolithography and lift-off. The heaters have a width of 6 \textmu m, and are sufficiently far from the waveguide layer so as not to introduce any optical loss due to interaction with the metal \cite{ji_breaking_2016}.

\subsection{Experimental setup}
A continuous wave (cw) RIO Orion laser at 1561.4 nm with a narrow linewidth of $<$ 3 kHz is used as the pump. It is amplified by an erbium doped fibre amplifier (EDFA) and filtered by a 27.5 GHz bandwidth DWDM add-drop filter centered at 1561.42 nm to mitigate the amplified spontaneous emission noise generated by the EDFA. A fibre polarization controller (FPC) before the chip and a polarizer after the chip are used to couple to the fundamental transverse electric (TE) mode of the bus waveguide using a lensed fibre. The waveguide is coupled to rings R1 and R2, which have platinum microheaters on them to enable thermal tuning of the rings' resonance wavelength using the thermo-optic effect. Light is coupled out of the chip using a 40x aspheric lens. The microheaters are driven using tungsten probes by an arbitrary waveform generator (AWG) in burst mode, with a staggered delay ($\sim$ 0.85 ms) between the burst triggers of the two voltages V1 and V2 applied to the heaters on rings R1 and R2 respectively. On coupling 110 mW of power into the bus waveguide and applying the requisite voltages to the rings (shown in Fig. \ref{fig:2_optical}(e)), soliton mode-locked dual combs are generated. A part of the output of the chip is sent to an optical spectrum analyzer (OSA) to monitor comb formation. The rest is sent to another DWDM add-drop filter to separate the pump and the dual comb. The pump and the generated dual comb are both sent to separate photodiodes and the dc component is monitored on an oscilloscope. The dual comb is also sent to a 45 GHz fast photodiode (Newport 1014). For dual-comb spectroscopy, a cuvette with a path length of 10 mm, containing the sample to be measured (dichloromethane in our experiment) is placed in the path of the dual comb. The RF output of the fast photodiode is amplified using a three stage broadband amplifier (Centellax OA4MVM3) with a bandwidth of 40 GHz and sent to a 36 GHz oscilloscope (LeCroy LabMaster 10 Zi-A, 8-bit resolution, 80 GSa/s). A time domain trace of the dual-comb multiheterodyne beating is acquired, followed by a fast Fourier transform (FFT) to obtain the RF beat notes. The fast oscilloscope can also be used as an RF spectrum analyzer for measuring the linewidth of the beat notes. A detailed outline of the experimental setup can be found in the supplementary information Fig. S2.

\section*{Acknowledgements}
We acknowledge fruitful discussions with Dr. Nathalie Piqu\'{e} from the Max-Planck Institute of Quantum Optics, Garching, Germany, and with Alexander Klenner, Gaurang Bhatt, Steven Miller, Aseema Mohanty and Mengjie Yu from Columbia University, for the DCS measurements. This work was performed in part at the Cornell NanoScale Facility, a member of the National Nanotechnology Coordinated Infrastructure (NNCI), which is supported by the National Science Foundation (NSF) (Grant ECCS-1542081). This work made use of the Cornell Center for Materials Research Shared Facilities which are supported through the NSF MRSEC program (DMR-1120296). The authors acknowledge support from the Defense Advanced Research Projects Agency (N66001-16-1-4052, W31P4Q-15-1-0015) and the Air Force Office of Scientific Research (FA9550-15-1-0303).

\bibliographystyle{naturemag}
\bibliography{library_journal_2016_10_18}

\begin{widetext}

\begin{figure}
\includegraphics[width=\textwidth]{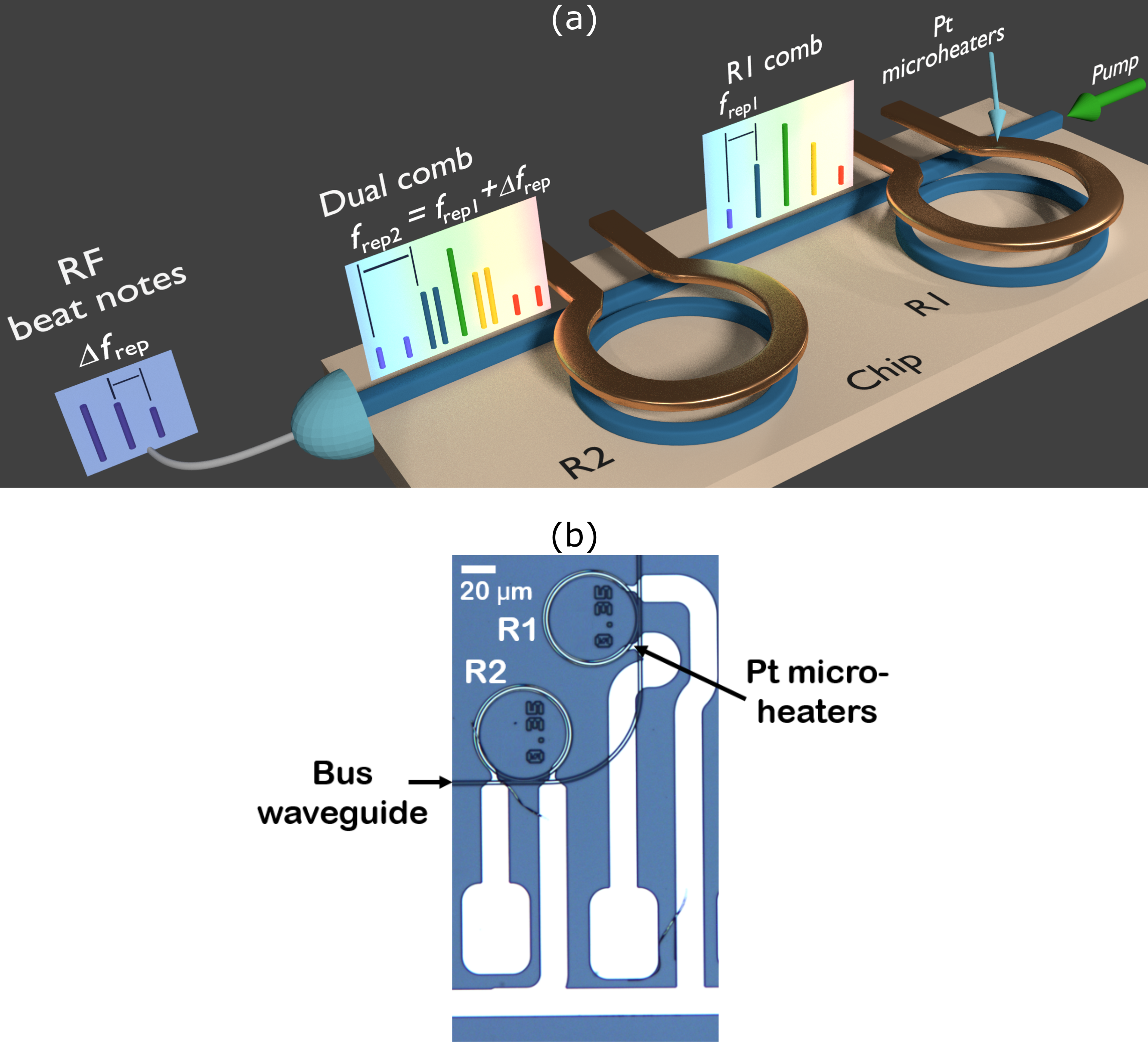}
\caption{{\bf Cascaded microring resonators for dual comb generation.} (a) Schematic of the device. A cw laser pumps the silicon nitride waveguide, which is coupled to two silicon nitride rings, R1 and R2. Through parametric oscillation and cascaded four-wave mixing, rings R1 and R2 generate frequency combs with repetition rates $f_{\rm rep1}$ and $f_{\rm rep2}$ respectively. The insets show schematic optical spectra after the first and second rings. The spectrum detected at the output of the chip using a fast photodiode shows a series of beat notes in the RF domain. Note that while the optical spectrum insets span several tens of THz, the RF spectrum spans only a few GHz enabling ease of detection in the electronic domain. (b) Optical microscope image of the fabricated device showing the silicon nitride rings with integrated platinum microheaters.}
\label{fig:1_schematic}
\end{figure}

\begin{figure}
\includegraphics[width=.85\textwidth]{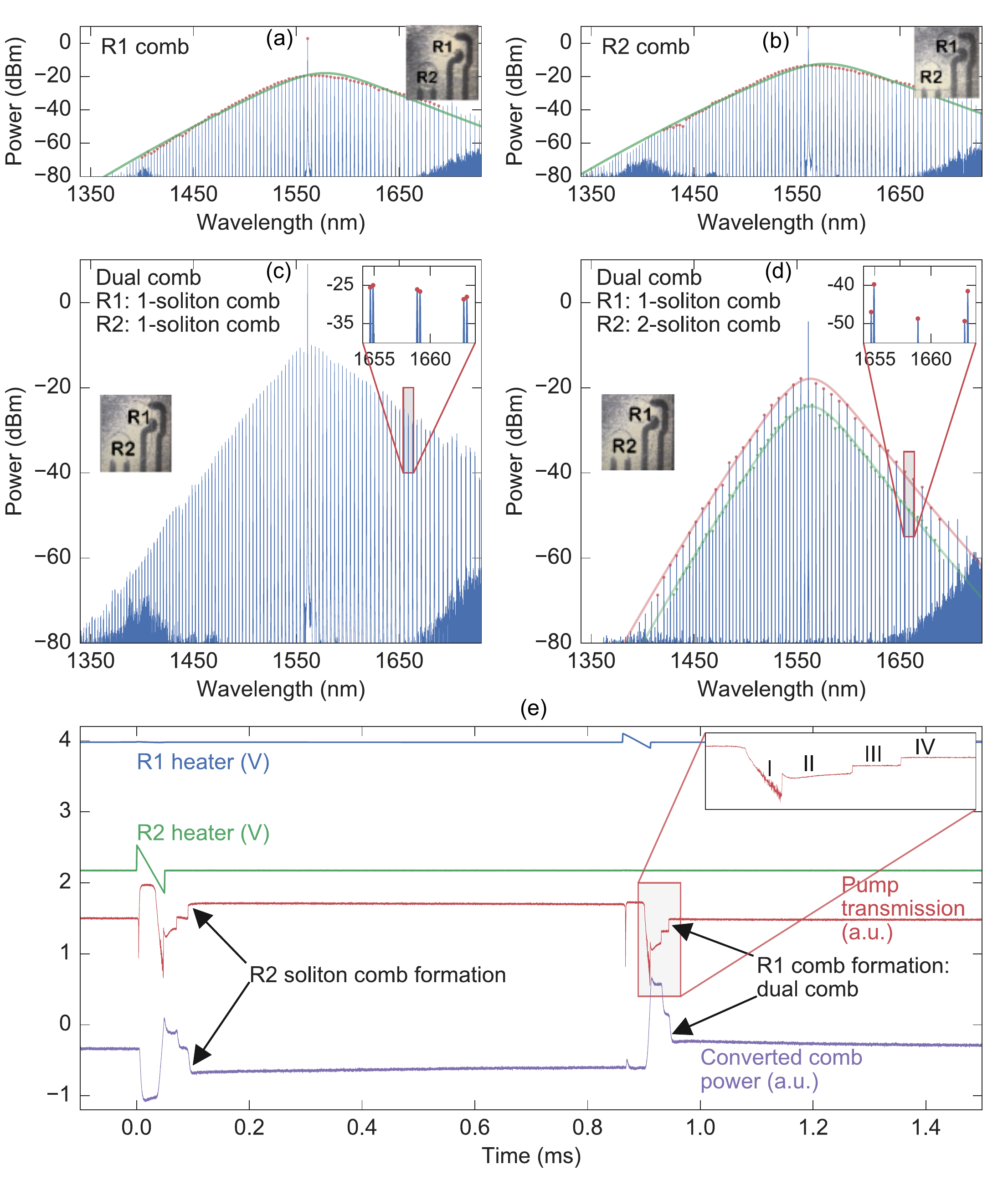}
\caption{\small {\bf Optical spectra of the generated combs.} {\bf (a)-(c)} Generated optical spectrum when the heaters on (a) R1, (b) R2 and (c) both rings are tuned to resonance with the laser. The spectra were acquired with an OSA with a resolution of 0.01 nm. The insets show top-view infrared CCD camera images showing which of the two rings are on resonance with the pump laser. The waveguide has been false coloured for clarity. The generated combs are in the single-soliton mode-locked state with smooth sech$^2$ spectral envelopes, shown by the solid green lines. The zoom-in inset in (c) and (d) also reveals the dual comb with pairs of lines closely spaced in wavelength. {\bf (d)} A dual comb state with a single soliton in R1 and two solitons in R2. By changing the magnitude of the final upward voltage ramp shown in (e), one can choose the number of solitons in the final mode-locked state. As an example, the state shown here has a single soliton in R1 and two solitons half a roundtrip apart in R2 (harmonic mode-locking). All spectra span 400 nm (51 THz) around the pump wavelength of 1561.4 nm. {\bf (e)} Time domain traces of the voltage ramp used to access the mode-locked states of both the rings. The blue (green) line represents the voltage applied to the heater on R1 (R2). The red and purple lines represent the transmitted pump power and the converted comb power (excluding the pump) respectively. Discrete steps are seen in the transmitted pump power as well as the converted comb power, characteristic of the formation of multi-soliton mode-locked states in the ring, eventually leading to the final single soliton state. The inset shows the distinct stages of comb formation in R1, including the high-noise state (I), the multi-soliton state (II and III) and the single soliton state (IV). Note that each voltage ramp on the heaters of R1 and R2 is 50 \textmu s long, corresponding to a tuning speed of 20 kHz. The converted comb power has been offset vertically for clarity.}
\label{fig:2_optical}
\end{figure}

\begin{figure}
\includegraphics[width=\textwidth]{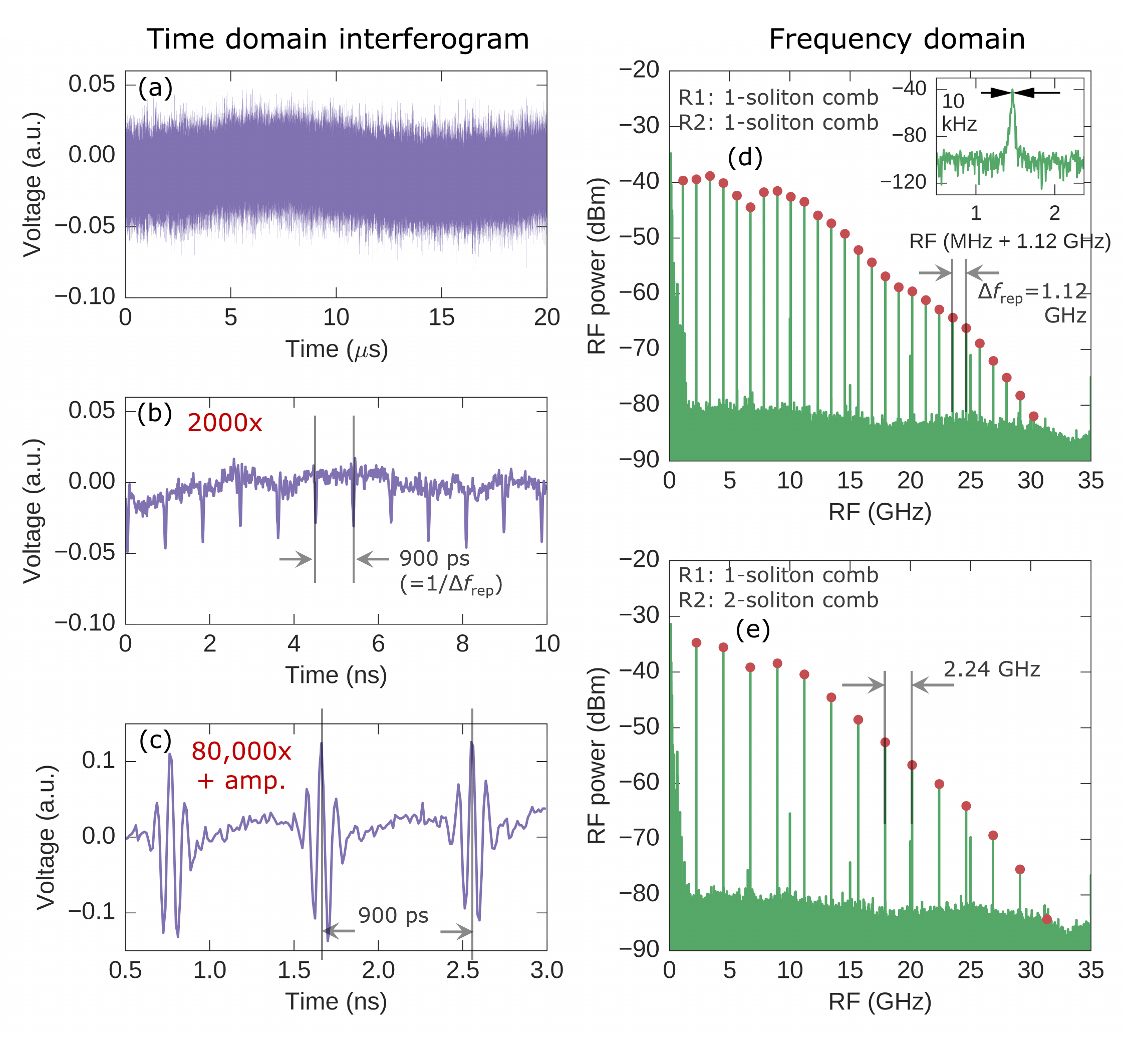}
\caption{{\bf Time domain interferogram and RF heterodyne beat note.} {\bf (a)} 20 \textmu s time domain interferogram of the mode-locked dual comb acquired with a fast photodiode and a 36 GHz oscilloscope. {\bf (b)} The same time domain interferogram expanded 2000 times shows periodic pulses at the inverse of the beat note frequency ($1/\dfr$). Note the different scale on the x-axis compared to (a). {\bf (c)} The dual comb is amplified by an L-band EDFA and detected on a fast photodiode to highlight features of the time domain interferogram. This trace is zoomed in by a factor of 80,000 compared to (a). {\bf (d)} RF multi-heterodyne beat notes obtained by performing a fast Fourier transform (FFT) on the interferogram in (a). The data reported here was taken with a long pass filter and represents the dual comb lines on the red side of the pump. Similar spectra were observed with a short pass filter for the blue side of the comb. The beat notes correspond to the dual comb shown in Fig. \ref{fig:2_optical}(c), where both combs are in the single-soliton mode-locked state. A beat note spacing of 1.12 GHz is observed. The inset shows the linewidth of the first beat note, measured using an RF spectrum analyzer with a resolution bandwidth of 2 kHz. We measure a high SNR (60 dB) with a linewidth of 10 kHz, which corresponds to a relative coherence time of 100 \textmu s. {\bf (e)} RF beat notes of the comb shown in Fig. \ref{fig:2_optical}(d), where the comb generated in R1 is in the single soliton mode-locked state and the comb in R2 is in the two-soliton harmonic mode-locked state. The beat notes are spaced by 2.24 GHz as the R2 comb is missing a line in every alternate mode of the ring. In both (d) and (e), the beat notes close to the pump have a high SNR in excess of 40 dB. The spurious peaks at 5, 10, 15, 20 and 25 GHz are artefacts of the FFT analyzer.}
\label{fig:3_RF}
\end{figure}

\begin{figure}
\includegraphics[width=\textwidth]{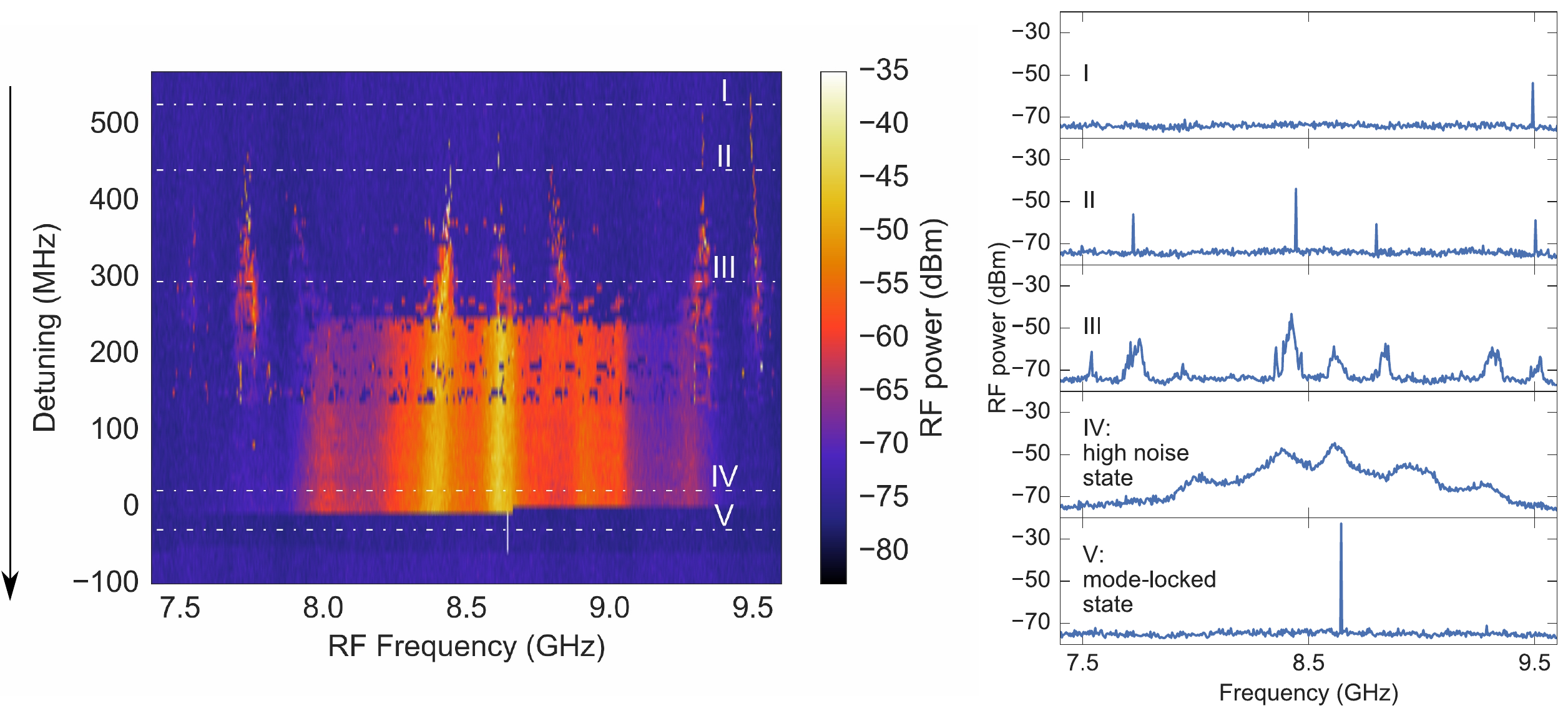}
\caption{{\bf Dual-comb heterodyne technique to study beat note evolution.} The device used in this measurement had a beat note spacing of 8.6 GHz. Keeping the comb generated in R1 in the mode-locked state, the resonance of R2 is tuned with respect to the pump laser and the RF spectrum is recorded. The left panel shows the spectrum of the first beat note as the heater power on R2 is varied starting from a far-blue-detuned state to the mode-locked state on the red detuned side. The spectra on the right show the formation of beat notes (Stage I and II), which broaden to form a wide beat note in the high noise state (Stage III and IV). On tuning further to the red-detuned side, we observe an abrupt transition to a narrow and high-SNR beat note as shown in Fig. \ref{fig:3_RF}. (see also Supplementary Movie and Supplementary Information section III).}
\label{fig:4_evolution}
\end{figure}

\begin{figure}
\includegraphics[width=\textwidth]{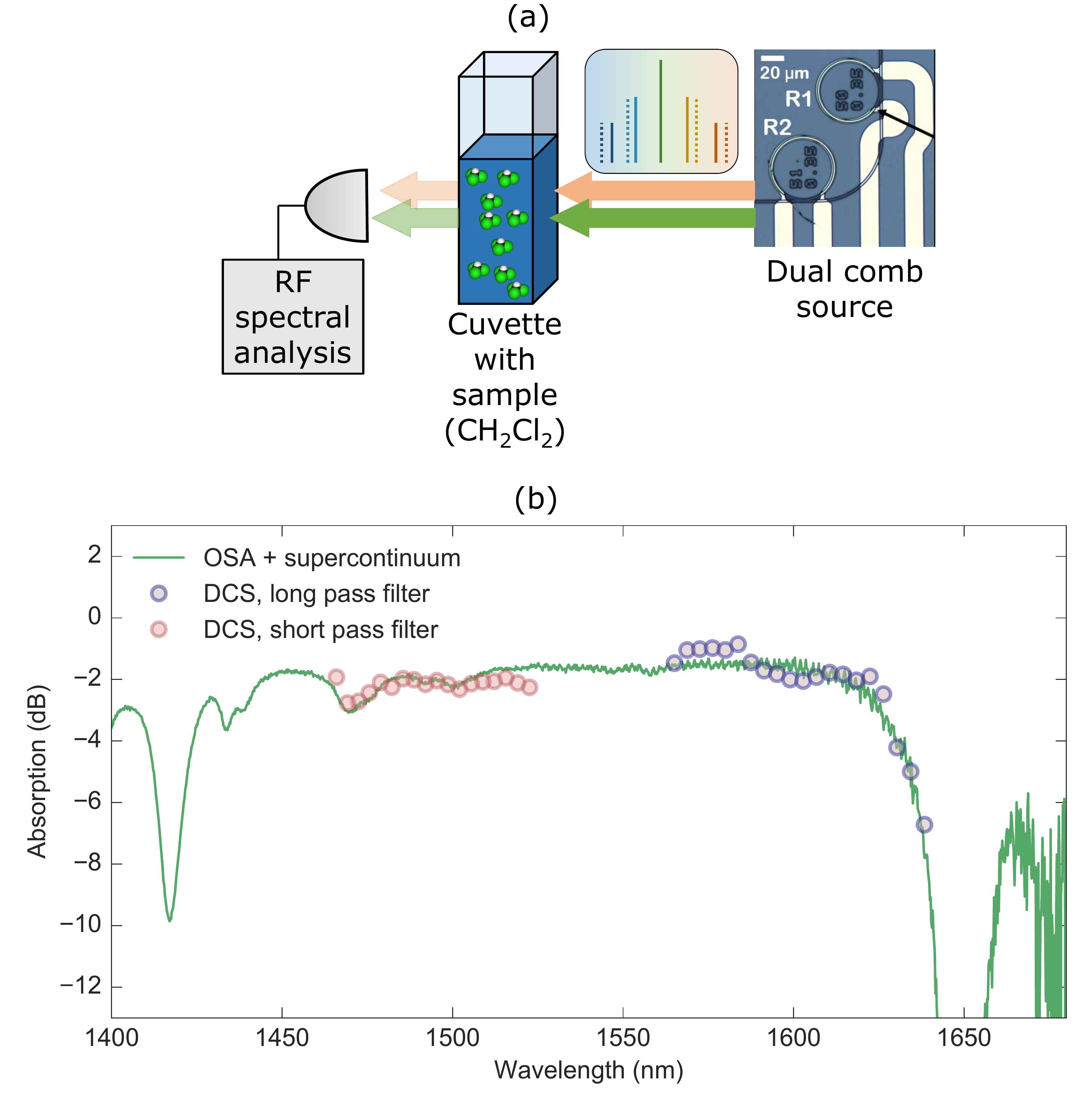}
\caption{{\bf Dual comb spectroscopy of dichloromethane.} {\bf (a)} Schematic of the setup. The output of the chip generating a dual comb is sent to a 10 mm cuvette containing dichloromethane, and the transmitted light is sent to a 45 GHz photodiode and analyzed WITH an RF spectrum analyzer or with a fast oscilloscope. {\bf (b)} Spectrum of dichloromethane acquired using DCS and corroborated with the same spectrum measured with a broadband supercontinuum source and an OSA. The solid line represents the absorption spectrum measured with a supercontinuum source (Fianium SC-450-4) and an OSA. The red (blue) circles represent the absorption spectrum acquired using the RF beat notes and a short pass (long pass) filter. All spectra are normalized by the corresponding spectrum without dichloromethane.}
\label{fig:5_DCS}
\end{figure}

\end{widetext}

\end{document}